| | |
|---|---|
| ***Numero*** | 68 |
| ***Anno di pubblicazione*** | 2020 |
| ***Data inserimento in OA@INAF*** | 2021-01-14T11:07:06Z |
| ***Titolo*** | Slaving and disabling actuators with voice-coil adaptive mirrors |
| ***Autori*** | Riccardi, A. |
| ***Afferenza primo autore*** | O.A. Arcetri |
| ***Handle*** | http://hdl.handle.net/20.500.12386/29767 |

INAF
TECHNICAL REPORT

Doc.No : 1/2012
Version : 3
Date : 06 Jul 2020

# Slaving and disabling actuators with voice-coil adaptive mirrors

| | |
|---|---|
| Prepared by | A. Riccardi |
| Approved by | A. Riccardi |
| Released by | A. Riccardi |

INAF – Osservatorio Astrofisico di Arcetri
Largo E. Fermi, 5 - 50125 Firenze - ITALY
**http://adopt.arcetri.astro.it**

# ABSTRACT

Adaptive mirrors based on voice-coil technology have force actuators with an internal metrology to close a local loop for controlling its shape in position. When actuators are requested to be disabled or slaved, control matrices have to be re-computed. The report describes the algorithms to re-compute the relevant matrixes for controlling of the mirror without the need of recalibration. This is related in particular to MMT, LBT, Magellan, VLT, ELT and GMT adaptive mirrors that use the voice-coil technology. The technique is successfully used in practice with LBT and VLT-UT4 adaptive secondary mirror units.

## Modification Record

| Version | Date | Author | Section/Paragraph affected | Reason/Remarks |
|---|---|---|---|---|
| 1 | 13 Jan 2012 | A. Riccardi | | First release of the document |
| 2 | 15 Apr 2012 | A. Riccardi | All | Cross references to Eqs. fixed |
| | | | Sec.3 | Eq.(5) and (7) missing term added |
| 3 | 06 Jul 2020 | A. Riccardi | All | Added minimum-RMS-force slaving algorithm |

## Abbreviations, acronyms and symbols

| Symbol | Description |
|---|---|
| ELT | Extremely Large Telescope |
| FF | Feed-Forward |
| GMT | Giant Magellan Telescope |
| IF | Interaction Function |
| IM | Interaction Matrix |
| LBT | Large Binocular Telescope |
| VLT | Very Large Telescope |

## Contents

# 1 Introduction

Adaptive mirrors based on voice-coil technology have force actuators with an internal metrology to close a local loop for controlling its shape in position. This technology is currently in use on 6-8m class telescope like LBT[1], Magellan[2] and VLT [3], and it will be part of Extremely Large Telescopes like the European ELT (M4)[4] and GMT (M2)[5].

When actuators are requested to be slaved or disabled, control matrices have to be re-computed in order to avoid a re-calibration of the system. A key matrix for the functioning of this kind of mirrors is the Feed-Forward (FF) matrix[6][7], i.e. the matrix relating the force requested to apply a given static shape. This matrix can be associated to the stiffness matrix of the deformable shell.[1] Let's distinguish two cases as following:

- **Disabling actuators**: this is the case when an actuator stops working for a failure of its force driver and its internal control loop is disabled. It cannot apply the FF force and the relative matrix has to be recomputed;
- **Slaving actuators**: Actuators outside the optical pupil having negligible interaction with the wavefront sensing cannot be disabled: they are clustered and disabling them would give a severe problem in mirror shell control. The control bandwidth is related to the damping per unit mass [1] and disabling actuators requires their mass to be controlled by the neighbors reducing their controllability. That is acceptable per sparse disabled actuators, but it is not acceptable for clustered ones. Those actuators (slaved actuators) require to be kept in local closed loop, so a position command is required for them. Because those have negligible sensitivity with the wavefront sensing of the adaptive optics loop, their position command has to be extrapolated from the commands of the other actuators (master actuators).

# 2 Relevant matrixes and their definitions

The matrixes that are involved in the re-computation are defined in the following sub-section.

## 2.1 The Feed-Forward matrix K

The FF matrix $\boldsymbol{K}$ relates the force vector $\boldsymbol{f}$ that is needed to apply a corresponding position command $\boldsymbol{p}$ as follows

(1) $\boldsymbol{f} = \boldsymbol{K}\boldsymbol{p}$ .

The FF matrix is defined for enabled actuators having both force driver and position sensor properly working.

## 2.2 The Influence Function Matrix M

The IF matrix $\boldsymbol{M}$ relates the position command vector $\boldsymbol{p}$ with the shape of the optical surface of the mirror $\boldsymbol{w}$ as follows

(2) $\boldsymbol{w} = \boldsymbol{M}\boldsymbol{p}$ .

## 2.3 The Interaction matrix D

The IM $\boldsymbol{D}$ relates the Wavefront Sensor signal vector $\boldsymbol{s}$ with mirror position vector $\boldsymbol{p}$ as follows

(3) $\boldsymbol{s} = \boldsymbol{D}\boldsymbol{p}$ .

# 3 Disabling actuators in the FF matrix

Let us consider the case a subset of actuators has to be disabled because no longer able to apply force or constrained for any reason to not apply force. In order to simplify notation, we will suppose the first $M$ actuators to keep in the list of enabled (master) actuators (identified by index $m$) and the last $S$ actuators to be disabled (identified by index[2] $s$). $N = M + S$ is the total number of degrees of freedom of the mirror. Eq. (1) can be re-written as

[1] The FF matrix is related to the stiffness matrix, but it is not the stiffness matrix. the latter is related to the displacement of the points of application of the forces applied to the shell, while FF matrix is related to the displacement as read by the internal position sensors that usually differ from the previous one both for geometry and calibration errors. In particular the stiffness matrix property to be invariant for transposition is not verified for of the FF matrix, i.e. $\boldsymbol{K}^T \neq \boldsymbol{K}$.

[2] In terms of algorithm, disabling is a particular case of slaving, as it will be shown later in the report. For this reason we are using here the index S (like Slaved) for disabled actuators to have a common notation throughout the report.

(4) $$\begin{pmatrix} \boldsymbol{f_m} \\ \boldsymbol{f_s} \end{pmatrix} = \begin{pmatrix} \boldsymbol{K_{mm}} & \boldsymbol{K_{ms}} \\ \boldsymbol{K_{sm}} & \boldsymbol{K_{ss}} \end{pmatrix} \begin{pmatrix} \boldsymbol{p_m} \\ \boldsymbol{p_s} \end{pmatrix} .$$

Because the disabled actuators do not apply force, for them $\boldsymbol{f_s} = 0$, while the corresponding position $\boldsymbol{p_s}$ is unknown because without control. Setting $\boldsymbol{f_s} = 0$ in Eq. (4) and solving for $f_m$ and $p_s$, we have

(5) $$\boldsymbol{f_m} = (\boldsymbol{K_{mm}} - \boldsymbol{K_{ms}}\, \boldsymbol{K_{ss}^{-1}} \boldsymbol{K_{sm}}) \boldsymbol{p_m}$$

(6) $$\boldsymbol{p_s} = -\boldsymbol{K_{ss}^{-1}} \boldsymbol{K_{sm}} \boldsymbol{p_m}$$

Eqs. (5) and (6) can be written in the following short from

(7) $$\boldsymbol{p_s} = \boldsymbol{Q_s}\, \boldsymbol{p_m}$$

(8) $$\boldsymbol{f_m} = \boldsymbol{K'}\, \boldsymbol{p_m}$$

when letting

(9) $$\boldsymbol{Q_s} = -\boldsymbol{K_{ss}^{-1}} \boldsymbol{K_{sm}}$$

as the position *slaving matrix* and

(10) $$\boldsymbol{K'} = \boldsymbol{K_{mm}} - \boldsymbol{K_{ms}} \boldsymbol{Q_s}$$

as the *reduced FF matrix*, i.e. the new FF matrix restricted to the set of master actuators.

# 4 Slaving actuators in the FF matrix

## 4.1 Zero-force slaving

The more straightforward way of slaving actuators is commanding them a set of positions to perform zero force. In this case, the slaved actuators command vector $p_s$ is given by Eq. (7) and (9) and the we will refer to this kind of slaving as *zero-force slaving*.

This slaving algorithm has the drawback to not allow the salved actuators to contribute the control of the shell shape at the border between salved and master actuators. In particular only master actuators are responsible of applying the correct derivative of the surface in the direction orthogonal to the border, inducing extra forces on the master actuators surrounding this area.

## 4.2 Minimum-rms-force slaving

To reduce the extra-force problem addressed in the previous section, the following alternative algorithm has been developed. The actuators are divided in three groups:

- A number $S$ of *slaved actuators*, identified with index $s$.
- A number $B$ of master actuators in a given area surrounding the border between master and slaved actuators. These actuators are identified with index $b$ and we will refer to them as *border-master actuators* or simply *border actuators*.
- The rest of the master actuators. These $I$ actuators are identified with index $i$ and we will refer to them as *inner-master actuators* or simply *inner actuators*.

$N = I + B + S$ is the total number of degrees of freedom of the mirror and $M = B + I$ is the total number of master actuators. The aim is to position the slaved actuators using a command that minimizes the RMS of forces of the joint set of slaved and border actuators. In this way the slaved actuators can drive forces helping the border actuators in keeping their commanded positions, but reducing their peak forces. We will refer to this algorithm as *minimum-RMS-force slaving*.

Eq. (1) can be re-written as

(11) $$\begin{pmatrix} \boldsymbol{f_i} \\ \boldsymbol{f_b} \\ \boldsymbol{f_s} \end{pmatrix} = \begin{pmatrix} \boldsymbol{K_{ii}} & \boldsymbol{K_{ib}} & \boldsymbol{K_{is}} \\ \boldsymbol{K_{bi}} & \boldsymbol{K_{bb}} & \boldsymbol{K_{bs}} \\ \boldsymbol{K_{si}} & \boldsymbol{K_{sb}} & \boldsymbol{K_{ss}} \end{pmatrix} \begin{pmatrix} \boldsymbol{p_i} \\ \boldsymbol{p_b} \\ \boldsymbol{p_s} \end{pmatrix}$$

and it is shown in Appendix 1 that the slaved actuator command $p_s$ to minimize the RMS forces can be expressed in the same short form as Eq. (7) as

(12) $\boldsymbol{p_s} = \boldsymbol{Q_s}\,\boldsymbol{p_m}$

where, this time, the vector of master actuator $\boldsymbol{p_m}$ and the slaving matrix $\boldsymbol{Q_s}$ are given by

(13) $\boldsymbol{p_m} = \begin{pmatrix}\boldsymbol{p_i}\\ \boldsymbol{p_b}\end{pmatrix}$

(14) $\boldsymbol{Q_s} = -\left(\boldsymbol{K_{bs}^T K_{bs}} + \boldsymbol{K_{ss}^T K_{ss}}\right)^{-1}\left(\boldsymbol{K_{bs}^T K_{bi}} + \boldsymbol{K_{ss}^T K_{si}} \quad \boldsymbol{K_{bs}^T K_{bb}} + \boldsymbol{K_{ss}^T K_{sb}}\right)$

Once the $\boldsymbol{p_s}$ vector is computed, the force vector for all the actuators is evaluated by Eq. (11).

In the limit case in which the force RMS is minimized on all the actuators, i.e. all the master actuators are border-master, Eq. (13) and (14) become

(15) $\boldsymbol{p_m} = \boldsymbol{p_b}$

(16) $\boldsymbol{Q_s} = -\left(\boldsymbol{K_{bs}^T K_{bs}} + \boldsymbol{K_{ss}^T K_{ss}}\right)^{-1}\left(\boldsymbol{K_{bs}^T K_{bb}} + \boldsymbol{K_{ss}^T K_{sb}}\right)$

### 4.3 Comparison between zero-force and minimum-RMS-force slaving algorithms

The comparison between the two algorithms is made here in the specific case of the Magellan Adaptive Secondary Mirror (ASM), but the conclusions can be easily generalized to any voice-coil DM requiring to slave a clustered set of actuators. In particular Magellan Telescope has a large central obscuration requiring to slave the 3 innermost ring of actuators. Figure 1 compares the forces applied by the actuators as a function of the radial distance for the focus and spherical aberration surface deformation when slaving the first 3 rings of actuators with the zero-force and the minimum-RMS-force algorithms. In this example the RMS is minimized over all the actuators. The reduction of the peak forces in the first 3 rings of master actuators surrounding the 3 central slaved rings is evident.

It can be shown that most of the peak force reduction is obtained by setting 1 ring of border-master actuators and there is no further reduction when 3 or more rings of border-master actuators are considered.

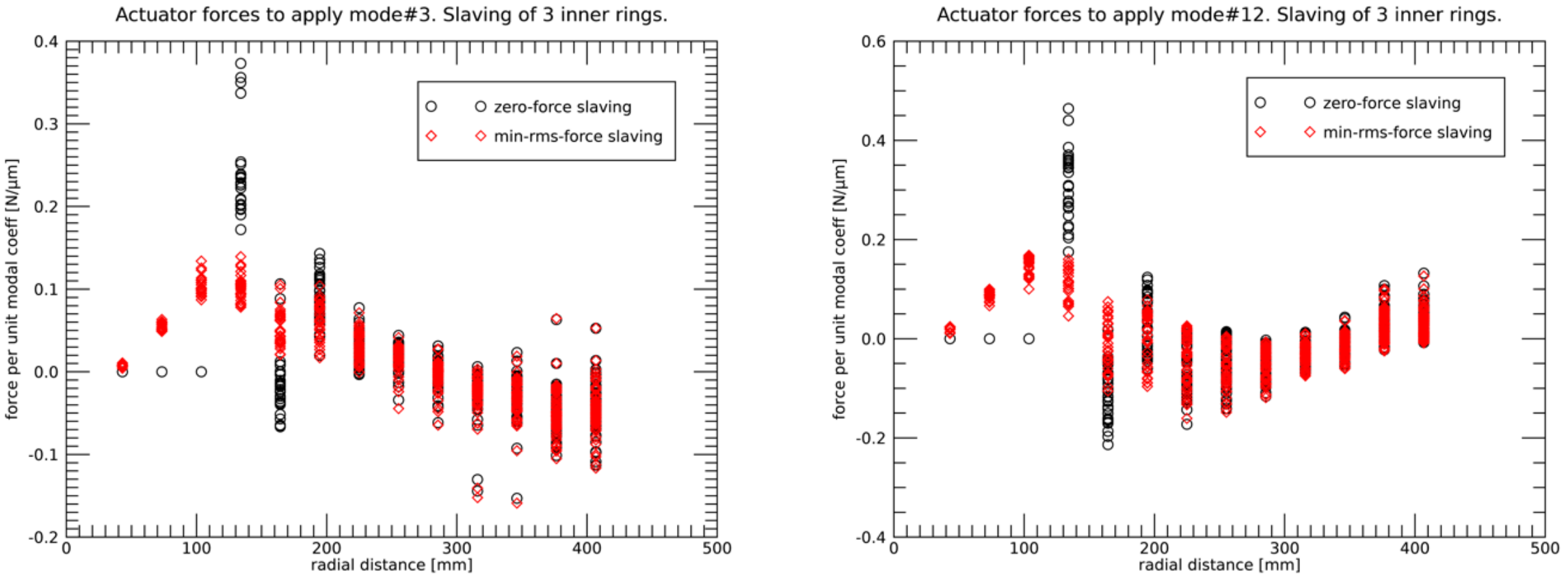

Figure 1 Comparison of actuators forces when a focus (on left) or a spherical aberration (on right) deformation is applied with the Magellan Adaptive Secondary Mirror using the zero-force and minimum-RMS-force slaving algorithms.

## 5 Influence Function matrix in case of disabled or slaved actuators

Defining

(17) $\boldsymbol{p} = \begin{pmatrix}\boldsymbol{p_m}\\ \boldsymbol{p_s}\end{pmatrix}$

and combining Eq. (2) with Eqs. (7) or (12) (depending on used algorithm), we have

(18) $\boldsymbol{w} = \boldsymbol{M}\boldsymbol{p} = (\boldsymbol{M}_m \quad \boldsymbol{M}_s)\begin{pmatrix}\boldsymbol{p}_m \\ \boldsymbol{p}_s\end{pmatrix} = (\boldsymbol{M}_m + \boldsymbol{M}_s\boldsymbol{Q}_s)\boldsymbol{p}_m = \boldsymbol{M}'\boldsymbol{p}_m$ ,

where

(19) $\boldsymbol{M}' = (\boldsymbol{M}_m - \boldsymbol{M}_s\boldsymbol{Q}_s)$ .

represents the new IF matrix restricted to the set of master actuators.

It has to be noted that in order to obtain the new IF matrix, the IFs of all the actuators are required. In case a modal IM has been calibrated with a number of modes that is less than the total number of actuators, that information is NOT sufficient to re-compute the IF matrix. An initial calibration of all $N$ zonal IFs or, equivalently, the full set of $N$ linearly independed modes is required, even if a limited number of modes is used for closing the optical loop.

## 6 Interaction Matrix in case of disabled or slaved actuators

Combining Eq. (3) and with Eqs. (7) or (12) (depending on used algorithm), we have

(20) $\boldsymbol{s} = \boldsymbol{D}\boldsymbol{p} = (\boldsymbol{D}_m \quad \boldsymbol{D}_s)\begin{pmatrix}\boldsymbol{p}_m \\ \boldsymbol{p}_s\end{pmatrix} = (\boldsymbol{D}_m - \boldsymbol{D}_s\boldsymbol{Q}_s)\boldsymbol{p}_m = \boldsymbol{D}'\boldsymbol{p}_m$ .

where

(21) $\boldsymbol{D}' = (\boldsymbol{D}_m - \boldsymbol{D}_s\boldsymbol{Q}_s)$ .

represents the new IM restricted to the set of master actuators.

## 7 Conclusions

The formulas to compute FF, IF and Interaction matrixes in case of slaved or disabled actuators have been computed, see Eqs. (8)-(10), (12)-(14), (19) and (21). The formulas for computing the position commands to send to the slaved actuators starting from the knowledge of the commands to the master actuators is also reported, see Eq. (7), (9), (14) and (16).

## Appendix 1

Considering Eq. (12), the RMS $\rho$ of the forces applied by the subset of slaved and border actuators is given by

(22) $\boldsymbol{\rho} = \boldsymbol{f}_b^T \boldsymbol{f}_b + \boldsymbol{f}_s^T \boldsymbol{f}_s = \boldsymbol{p}^T \boldsymbol{L}^T \boldsymbol{L} \boldsymbol{p}$

where

(23) $\boldsymbol{L} = \begin{pmatrix} \boldsymbol{K}_{bi} & \boldsymbol{K}_{bb} & \boldsymbol{K}_{bs} \\ \boldsymbol{K}_{si} & \boldsymbol{K}_{sb} & \boldsymbol{K}_{ss} \end{pmatrix}$

and

(24) $\boldsymbol{p} = \begin{pmatrix} \boldsymbol{p}_i \\ \boldsymbol{p}_b \\ \boldsymbol{p}_s \end{pmatrix}$

The force RMS in Eq. (22) has a minimum value when its differential $\mathbf{d\rho}$ is zero

(25) $\boldsymbol{d\rho} = \boldsymbol{d}(\boldsymbol{p}^T \boldsymbol{L}^T \boldsymbol{L} \boldsymbol{p}) = \boldsymbol{dp}^T \boldsymbol{L}^T \boldsymbol{L} \boldsymbol{p} + \boldsymbol{p}^T \boldsymbol{L}^T \boldsymbol{L}\, \boldsymbol{dp} = \boldsymbol{dp}^T \boldsymbol{L}^T \boldsymbol{L} \boldsymbol{p} + (\boldsymbol{dp}^T \boldsymbol{L}^T \boldsymbol{L} \boldsymbol{p})^T = \mathbf{2}\, \boldsymbol{dp}^T \boldsymbol{L}^T \boldsymbol{L} \boldsymbol{p} = \mathbf{0}$

where $(\boldsymbol{d}\mathrm{p}^T \mathrm{L}^T \mathrm{Lp})^T = \boldsymbol{d}\mathrm{p}^T \mathrm{L}^T \mathrm{Lp}$ because $\boldsymbol{d}\mathrm{p}^T \mathrm{L}^T \mathrm{Lp}$ is a scalar.

The positions of the master actuators $\boldsymbol{p}_i$ and $\boldsymbol{p}_b$ are constrained to be fixed, while the positions $\boldsymbol{p}_s$ of the slaved actuators can be changed to minimize $\boldsymbol{\rho}$, then the differential position dp has the form

(26) $\boldsymbol{dp} = \begin{pmatrix} \mathbf{0} \\ \mathbf{0} \\ \boldsymbol{dp}_s \end{pmatrix}$

and then from Eq. (25)

(27) $(\mathbf{0} \quad \mathbf{0} \quad \boldsymbol{dp}_s^T) \begin{pmatrix} \boldsymbol{K}_{bi}^T & \boldsymbol{K}_{si}^T \\ \boldsymbol{K}_{bb}^T & \boldsymbol{K}_{sb}^T \\ \boldsymbol{K}_{bs}^T & \boldsymbol{K}_{ss}^T \end{pmatrix} \begin{pmatrix} \boldsymbol{K}_{bi} & \boldsymbol{K}_{bb} & \boldsymbol{K}_{bs} \\ \boldsymbol{K}_{si} & \boldsymbol{K}_{sb} & \boldsymbol{K}_{ss} \end{pmatrix} \begin{pmatrix} \boldsymbol{p}_i \\ \boldsymbol{p}_b \\ \boldsymbol{p}_s \end{pmatrix} = \mathbf{0}$

(28) $\boldsymbol{dp}_s^T (\boldsymbol{K}_{bs}^T \quad \boldsymbol{K}_{ss}^T) \begin{pmatrix} \boldsymbol{K}_{bi} & \boldsymbol{K}_{bb} & \boldsymbol{K}_{bs} \\ \boldsymbol{K}_{si} & \boldsymbol{K}_{sb} & \boldsymbol{K}_{ss} \end{pmatrix} \begin{pmatrix} \boldsymbol{p}_i \\ \boldsymbol{p}_b \\ \boldsymbol{p}_s \end{pmatrix} = \mathbf{0}$

Because the above equation must be valid for any set of $\mathrm{d}\boldsymbol{p}_s^T$,

(29) $(\boldsymbol{K}_{bs}^T \quad \boldsymbol{K}_{ss}^T) \begin{pmatrix} \boldsymbol{K}_{bi} & \boldsymbol{K}_{bb} & \boldsymbol{K}_{bs} \\ \boldsymbol{K}_{si} & \boldsymbol{K}_{sb} & \boldsymbol{K}_{ss} \end{pmatrix} \begin{pmatrix} \boldsymbol{p}_i \\ \boldsymbol{p}_b \\ \boldsymbol{p}_s \end{pmatrix} = \mathbf{0}$

Expanding the products we have

(30) $(\boldsymbol{K}_{bs}^T \boldsymbol{K}_{bi} + \boldsymbol{K}_{ss}^T \boldsymbol{K}_{si} \quad \boldsymbol{K}_{bs}^T \boldsymbol{K}_{bb} + \boldsymbol{K}_{ss}^T \boldsymbol{K}_{sb} \quad \boldsymbol{K}_{bs}^T \boldsymbol{K}_{bs} + \boldsymbol{K}_{ss}^T \boldsymbol{K}_{ss}) \begin{pmatrix} \boldsymbol{p}_i \\ \boldsymbol{p}_b \\ \boldsymbol{p}_s \end{pmatrix} = \mathbf{0}$

(31) $(\boldsymbol{K}_{bs}^T \boldsymbol{K}_{bi} + \boldsymbol{K}_{ss}^T \boldsymbol{K}_{si}) \boldsymbol{p}_i + (\boldsymbol{K}_{bs}^T \boldsymbol{K}_{bb} + \boldsymbol{K}_{ss}^T \boldsymbol{K}_{sb}) \boldsymbol{p}_b + (\boldsymbol{K}_{bs}^T \boldsymbol{K}_{bs} + \boldsymbol{K}_{ss}^T \boldsymbol{K}_{ss}) \boldsymbol{p}_s = \mathbf{0}$

Solving for $\mathrm{p}_s$

(32) $(\boldsymbol{K}_{bs}^T \boldsymbol{K}_{bs} + \boldsymbol{K}_{ss}^T \boldsymbol{K}_{ss}) \boldsymbol{p}_s = -(\boldsymbol{K}_{bs}^T \boldsymbol{K}_{bi} + \boldsymbol{K}_{ss}^T \boldsymbol{K}_{si}) \boldsymbol{p}_i - (\boldsymbol{K}_{bs}^T \boldsymbol{K}_{bb} + \boldsymbol{K}_{ss}^T \boldsymbol{K}_{sb}) \boldsymbol{p}_b$

(33) $(\boldsymbol{K}_{bs}^T \boldsymbol{K}_{bs} + \boldsymbol{K}_{ss}^T \boldsymbol{K}_{ss}) \boldsymbol{p}_s = -(\boldsymbol{K}_{bs}^T \boldsymbol{K}_{bi} + \boldsymbol{K}_{ss}^T \boldsymbol{K}_{si} \quad \boldsymbol{K}_{bs}^T \boldsymbol{K}_{bb} + \boldsymbol{K}_{ss}^T \boldsymbol{K}_{sb}) \begin{pmatrix} \boldsymbol{p}_i \\ \boldsymbol{p}_b \end{pmatrix}$

(34) $\boldsymbol{p}_s = -(\boldsymbol{K}_{bs}^T \boldsymbol{K}_{bs} + \boldsymbol{K}_{ss}^T \boldsymbol{K}_{ss})^{-1} (\boldsymbol{K}_{bs}^T \boldsymbol{K}_{bi} + \boldsymbol{K}_{ss}^T \boldsymbol{K}_{si} \quad \boldsymbol{K}_{bs}^T \boldsymbol{K}_{bb} + \boldsymbol{K}_{ss}^T \boldsymbol{K}_{sb}) \begin{pmatrix} \boldsymbol{p}_i \\ \boldsymbol{p}_b \end{pmatrix}$

Setting

(35) $\boldsymbol{Q_s} = -\left(\boldsymbol{K_{bs}^T K_{bs}} + \boldsymbol{K_{ss}^T K_{ss}}\right)^{-1}\left(\boldsymbol{K_{bs}^T K_{bi}} + \boldsymbol{K_{ss}^T K_{si}} \quad \boldsymbol{K_{bs}^T K_{bb}} + \boldsymbol{K_{ss}^T K_{sb}}\right)$

(36) $\boldsymbol{p_m} = \begin{pmatrix} \boldsymbol{p_i} \\ \boldsymbol{p_b} \end{pmatrix}$

Eq. (34) can be written in the following short form

(37) $\boldsymbol{p_s} = \boldsymbol{Q_s}\, \boldsymbol{p_m}$

## Appendix 2

Substituting Eq. (37) in Eq. (11) and performing some algebra, the relation between the force vector $\boldsymbol{f}$ for all the actuators and the position vector $\boldsymbol{p_m}$ of the master actuators can be written in the short form as

(38) $\boldsymbol{f} = \boldsymbol{K'}\, \boldsymbol{p_m}$

where

(39) $\boldsymbol{f} = \begin{pmatrix} \boldsymbol{f_i} \\ \boldsymbol{f_b} \\ \boldsymbol{f_s} \end{pmatrix}$

and

(40) $\boldsymbol{K'} = \begin{pmatrix} \boldsymbol{K_{ii}} & \boldsymbol{K_{ib}} \\ \boldsymbol{K_{bi}} & \boldsymbol{K_{bb}} \\ \boldsymbol{K_{si}} & \boldsymbol{K_{sb}} \end{pmatrix} + \begin{pmatrix} \boldsymbol{K_{is}} \\ \boldsymbol{K_{bs}} \\ \boldsymbol{K_{ss}} \end{pmatrix} \boldsymbol{Q_s}$

## References

[1] A. Riccardi, M. Xompero, R. Briguglio, F. Quirós-Pacheco, L. Busoni, L. Fini, A. Puglisi, S. Esposito, C. Arcidiacono, E. Pinna, P. Ranfagni, P. Salinari, G. Brusa, R. Demers, R. Biasi, and D. Gallieni, *"The adaptive secondary mirror for the Large Binocular Telescope: optical acceptance test and preliminary on-sky commissioning results,"* in Society of Photo-Optical Instrumentation Engineers (SPIE) Conference Series, vol. 7736 of Presented at the Society of Photo-Optical Instrumentation Engineers (SPIE) Conference, July 2010 http://spiedigitallibrary.org/proceedings/resource/2/psisdg/7736/1/77362C_1

[2] Close, L. M. *et al*. First closed-loop visible AO test results for the advanced adaptive secondary AO system for the Magellan Telescope: MagAO's performance and status. In *Adaptive Optics Systems III*, vol. 8447 of *Proc. of SPIE*, 84470X (2012) https://www.spiedigitallibrary.org/conference-proceedings-of-spie/8447/1/First-closed-loop-visible-AO-test-results-for-the-advanced/10.1117/12.926545.short

[3] R. Arsenault, R. Biasi, D. Gallieni, A. Riccardi, P. Lazzarini, N. Hubin, E. Fedrigo, R. Donaldson, S. Oberti, S. Stroebele, R. Conzelmann, and M. Duchateau, *"A deformable secondary mirror for the VLT,"* in Advances in Adaptive Optics II, B. L. Ellerbroek and D. Bonaccini Calia, eds., vol. 6272 of Proc. SPIE, p. 0V, July 2006 http://adopt.arcetri.astro.it/html/pubblicazioni/arcetri/html/publications/2006_arsenault_spie_6272_0V.pdf

[4] D. Gallieni, M. Tintori, M. Mantegazza, E. Anaclerio, L. Crimella, M. Acerboni, R. Biasi, G. Angerer, M. Andrigettoni, A. Merler, D. Veronese, J. Carel, G. Marque, E. Molinari, D. Tresoldi, G. Toso, P. Spanó, M. Riva, R. Mazzoleni, A. Riccardi, P. Mantegazza, M. Manetti, M. Morandini, E. Vernet, N. Hubin, L. Jochum, P. Madec, M. Dimmler, and F. Koch, "Voice-coil technology for the E-ELT M4 Adaptive Unit," in Adaptative Optics for Extremely Large Telescopes, 2010. http://ao4elt.edpsciences.org/articles/ao4elt/pdf/2010/01/ao4elt_06002.pdf

[5] R. Biasi, D. Veronese, M. Andrighettoni, G. Angerer, D. Gallieni, M. Mantegazza, M. Tintori, and P. Lazzarini, M. Manetti, M. W. Johns, P. M. Hinz, and J. Kern, "GMT adaptive secondary design," Proc. SPIE 7736, pag. 77363O (2010) http://spiedigitallibrary.org/proceedings/resource/2/psisdg/7736/1/77363O_1

[6] A. Riccardi, G. Brusa, C. Del Vecchio, P. Salinari, R. Biasi, M. Andrighettoni, D. Gallieni, F. Zocchi, M. Lloyd-Hart, F. Wildi, and H. M. Martin, *"The adaptive secondary mirror for the 6.5 conversion of the Multiple Mirror Telescope,"* in Beyond Conventional Adaptive Optics, vol. 58 of ESO Proc., pp. 55-64, 2001 http://adopt.arcetri.astro.it/html/pubblicazioni/arcetri/html/publications/2001_riccardi_eso_58_55.pdf

[7] G. Brusa, A. Riccardi, M. Accardo, V. Biliotti, M. Carbillet, C. Del Vecchio, S. Esposito, B. Femenía, O. Feeney, L. Fini, S. Gennari, L. Miglietta, P. Salinari, and P. Stefanini, *" From adaptive secondary mirrors to extra-thin extra-large adaptive primary mirrors,"* in Proceedings of the Backaskog workshop on extremely large telescopes, T. Andersen, A. Ardeberg, and R. Gilmozzi, eds., vol. 57 of ESO Proc., pp. 181-201, Lund Obs. and ESO, 2000 http://adopt.arcetri.astro.it/html/pubblicazioni/arcetri/html/publications/1999_brusa_eso_57_181.pdf

**END OF DOCUMENT**